\documentclass[12pt]{article}
\usepackage{amssymb,amsmath,epsfig}

\begin{document}
\title{\bf Structure Scalars in Charged Plane Symmetry}

\author{M. Sharif \thanks{msharif.math@pu.edu.pk} and M. Zaeem Ul Haq Bhatti
\thanks{mzaeem.math@gmail.com}\\
Department of Mathematics, University of the Punjab,\\
Quaid-e-Azam Campus, Lahore-54590, Pakistan.}

\date{}

\maketitle
\begin{abstract}
We consider non-adiabatic flow of the fluid possessing dissipation
in the form of shearing viscosity in electromagnetic field. The
scalar functions (structure scalars) for charged plane symmetry are
formulated and are related with the physical variables of the fluid.
We also develop a relationship between the Weyl tensor and other
physical variables by using Taub mass formalism. The role of
electric charge as well as its physical significance for the
evolution of the shear tensor and expansion scalar are also
explored. Finally, we discuss a special case for dust with
cosmological constant.
\end{abstract}

\section{Introduction}

Self-gravitation is a process through which different components of
a large body are binded together. In spite of it, stars, stellar
clusters, galaxies and clusters of galaxies would all expand and
dissipate. Self-gravitating fluids have attracted many people due to
their applications in general relativity, astrophysics and
cosmology. These are usually characterized by the set of physical
variables. Many structures have been identified which produce
pressure anisotropy in stellar models and make the fluid imperfect.
The effect of local anisotropy is usually taken into consideration
under static conditions \cite{1}.

The inclusion of an electromagnetic field has shown interesting
outcome in gravitational collapse. Bekenstein \cite{50} was the
first who explored the Oppenheimer-Volkoff equations of hydrostatic
equilibrium \cite{51} from the neutral to the charged case. In the
discussion of charged gravitational collapse, Nath et al. \cite{52}
concluded that electromagnetic field increases the formation of
naked singularity. Sharif and Abbas \cite{53a} analyzed the effect
of electromagnetic field on gravitational collapse with positive
cosmological constant and found two apparent horizons (cosmological
and black hole) whose area decreases in the presence of
electromagnetic field. Further, they \cite{53b} extended this study
from four dimensions to five dimensions in the presence of
electromagnetic field and examined that the electromagnetic field
reduces the pressure but favors the formation of naked singularity.

The same authors \cite{53c} discussed the gravitational collapse for
cylindrical symmetry with charged perfect fluid. Sharif and Siddiqa
\cite{53d} carried out the consequences of charge and dissipation
for plane symmetrical gravitational collapse of real fluid. The
dynamical and transport equations are coupled to check the effects
of dissipation over collapsing process. Sharif and Fatima \cite{53e}
discussed the coupled dynamical and transport equation for
cylindrically symmetric collapsing process. Moreover, they
formulated a relationship between the Weyl tensor and energy-density
inhomogeneity. Rosales et al. \cite{4} pointed out that electric
charge plays the same role of anisotropy in the collapse \cite{5},
when the tangential pressure is greater than radial pressure.

Herrera et al. \cite{11a} figured out a systematic study of
spherically symmetric self-gravitating relativistic fluids, based on
the scalar functions (structure scalars) derived from the orthogonal
splitting of the Riemann tensor. Also, Herrera et al. \cite{11b}
provided a study on two aspects of Lemaitre-Tolman-Bondi (LTB)
spacetimes and gave an alternatives for obtaining spherically
symmetric dust solutions. Moreover, Herrera et al. \cite{11c}
investigated the stability of shear free condition based on the
evolution equation of the shear tensor and found that the major role
is played by the scalar $Y_{TF}$. Recently, Herrera et al.
\cite{11d} provided a detailed study based on the structure and
evolution of self-gravitating relativistic fluids through structure
scalars. These functions are based on the orthogonal splitting
\cite{12} of the Riemann tensor in general relativity which are
denoted by $X_T, X_{TF}, Y_T, Y_{TF}$. The role of electric charge
and cosmological constant on structure scalars is also analyzed for
spherically symmetric spacetime \cite{13}. In a recent paper, we
have investigated the structure scalars for the charged
cylindrically symmetric spacetime \cite{13a}.

This paper develops structure scalars for the charged plane
symmetric spacetime. The format is as follows. Section \textbf{2} is
devoted for the discussion of fluid distribution and formulate the
Einstein-Maxwell field equations. We also find a relation between
the Weyl tensor and energy density inhomogeneity. Section \textbf{3}
contains the discussion of structure scalars. In section \textbf{4},
we establish the structure scalars for the dust case with
cosmological constant in the absence of dissipation and viscosity.
Finally, we summarize the results in the last section.

\section{Fluid Distribution and the Field Equations}

We consider a general non-rotating plane symmetric distribution of
collapsing fluid whose line element is given by
\begin{equation}\label{1}
ds^2=-A^2(t,z)dt^{2}+B^2(t,z)\left(dx^{2}+dy^2\right)+C^2(t,z)dz^2.
\end{equation}
We assume a fluid distribution which is locally anisotropic and
suffering dissipation in the form of shearing viscosity, heat flow
and free streaming radiation. Its energy momentum-tensor is
\begin{equation}\label{2}
T_{\alpha\beta}=(\mu+P_{\bot})V_\alpha V_\beta+P_\bot
g_{\alpha\beta} + (P_z - P_\bot)\chi_\alpha\chi_\beta+q_\alpha
V_\beta+ V_\alpha q_\beta+{\epsilon}l_\alpha
l_\beta-2{\eta}{\sigma}_{\alpha\beta},
\end{equation}
where $\mu,~P_{\perp},~V^{\alpha},~q_{\alpha},~\epsilon,~\eta$ and
$\sigma_{\alpha\beta}$ are the energy density, tangential pressure,
four velocity, heat flux, radiation density, coefficient of shear
viscosity and the shear tensor, respectively. Also,
$P_z,~\chi^\alpha$ and $l^\alpha$ are pressure, unit four-vector and
the null four-vector in the $z$-direction, respectively. We assume
the fluid to be comoving so that Eq.(\ref{1}) satisfies
\begin{equation}\label{3}
V^{\alpha}=A^{-1}\delta^{\alpha}_{0},\quad
\chi^{\alpha}=C^{-1}\delta^{\alpha}_{3},\quad
q^\alpha=qC^{-1}\delta^{\alpha}_{3}, \quad
l^\alpha=A^{-1}\delta^{\alpha}_{0}+C^{-1}\delta^{\alpha}_{3},
\end{equation}
where $q$ is a function of $t$ and $z$ and
$q^{\alpha}=q\chi^\alpha$. The above quantities satisfy the
following relations
\begin{eqnarray}\nonumber
&&V^{\alpha}V_{\alpha}=-1,\quad\chi^{\alpha}\chi_{\alpha}=1,
\quad\chi^{\alpha}V_{\alpha}=0,\\\label{4}
&&V^\alpha q_\alpha=0, \quad l^\alpha V_\alpha=-1, \quad l^\alpha
l_\alpha=0.
\end{eqnarray}

The energy-momentum tensor in electromagnetic field is
\begin{equation}\label{5}
S_{\alpha\beta}=\frac{1}{4\pi}\left(F^{\gamma}_{\alpha}F_{\beta\gamma}-
\frac{1}{4}F^{\gamma\delta}F_{\gamma\delta}g_{\alpha\beta}\right),
\end{equation}
where $F_{\alpha\beta}=\phi_{\beta,\alpha}-\phi_{\alpha,\beta}$ is
the Maxwell field tensor and $\phi_\alpha$ is the four potential.
The Maxwell field equations are
\begin{equation}\label{6}
F^{\alpha\beta}_{;\beta}={\mu}_{0}J^{\alpha},\quad
F_{[\alpha\beta;\gamma]}=0,
\end{equation}
here $\mu_0=4\pi$ is the magnetic permeability and $J_\alpha$ is the
four current. In comoving coordinates, we have
\begin{equation*}
\phi_{\alpha}={\phi} {\delta^{\alpha}_{0}},\quad
J^{\alpha}={\xi}V^{\alpha},
\end{equation*}
where $\phi,~\xi$ represent the scalar potential and the charge
density, respectively, both are functions of $t$ and $z$. The charge
conservation equation, $J^\alpha_{;\alpha}=0$, yields
\begin{equation*}
s(z)=\int^z_{0}{\xi}C{B^2}dz.
\end{equation*}
Equation (\ref{6}) for $\alpha=0$ with ${\phi}'(t,0)=0$ gives
\begin{equation}\label{7}
{\phi}'=\frac{{\mu}_{0}s(z)AC}{B^{2}}.
\end{equation}
The Einstein field equations in electromagnetic field lead to
\begin{eqnarray}\nonumber
&&8{\pi}({\mu+\epsilon})A^{2}+\left(\frac{{\mu}_{0}sA}{B^{2}}\right)^2
=\frac{\dot{B}}{B}\left(\frac{2\dot{C}}{C}+\frac{\dot{B}}{B}\right)
-\left(\frac{A}{C}\right)^2\left[\frac{2B''}{B}\right.\\\label{8}
&&-\left(\frac{2C'}{C}\right.\left.-\frac{B'}{B}\right)\left.\frac{B'}{B}\right],\\\label{9}
&&-8{\pi}({q+\epsilon})AC=-2\left(\frac{\dot{B'}}{B}-\frac{A'\dot{B}}{AB}
-\frac{B'\dot{C}}{BC}\right),\\\nonumber
&&8{\pi}(P_{\bot}+\frac{2}{3}{\eta}F)B^{2}
+\left(\frac{{\mu}_{0}s}{B}\right)^{2}=-\left(\frac{B}{A}\right)^2\left[\frac{\ddot{B}}{B}\right.
-\frac{\ddot{C}}{C}
-\frac{\dot{A}}{A}\left(\frac{\dot{B}}{B}\right.\\\label{10}
&&+\left.\frac{\dot{C}}{C}\right)+\left.\frac{\dot{B}\dot{C}}{BC}\right]
+\left(\frac{B}{C}\right)^2\left[\frac{A''}{A}\right.
+\frac{B''}{B}-\frac{A'}{A}\left(\frac{C'}{C}-\frac{B'}{B}\right)
-\left.\frac{B'C'}{BC}\right],
\end{eqnarray}
\begin{eqnarray}\nonumber
&&8{\pi}(P_{z}+\epsilon-\frac{4}{3}{\eta}F)C^{2}-\left(\frac{{\mu}_{0}sC}{B^2}\right)^{2}
=-\left(\frac{C}{A}\right)^2\left[\frac{2\ddot{B}}{B}\right.
+\left(\frac{\dot{B}}{B}\right)^{2}\\\label{11}
&&-\left.\frac{2\dot{A}\dot{B}}{AB}\right]+\left(\frac{B'}{B}\right)^2+\frac{2A'B'}{AB},
\end{eqnarray}
where dot and prime stand for differentiation with respect to $t$
and $z$, respectively.

We can describe the non-rotating fluid by the kinematical variables,
i.e., expansion, acceleration and shear as
\begin{eqnarray}\nonumber
\Theta=V^{\alpha}_{~;\alpha},\quad
a_\alpha=V_{\alpha;\beta}V^\beta,\quad
\sigma_{\alpha\beta}=V_{(\alpha;\beta)}+a_{(\alpha}V_{\beta)}
-\frac{1}{3}\Theta h_{\alpha\beta},
\end{eqnarray}
where $h_{\alpha\beta}=g_{\alpha\beta}+V_{\alpha}V_\beta$ is the
projection tensor. Using Eq.(\ref{1}), these quantities turn out to
be
\begin{eqnarray}\label{12}
\Theta=\frac{1}{A}\left(2\frac{\dot{B}}{B}
+\frac{\dot{C}}{C}\right),\quad a_{3}=\frac{A'}{A},\quad a^{\alpha}
=a\chi^{\alpha},\quad a^2=a^\alpha a_{\alpha} =(\frac{A'}{AC})^2.
\end{eqnarray}
The magnitude of the shear tensor is defined as
\begin{equation*}
\sigma^2=\frac{1}{2}\sigma^{\alpha\beta}
\sigma_{\alpha\beta}=\frac{1}{9}F^2,
\end{equation*}
where
\begin{equation}\label{13}
F=\frac{1}{A}\left(-\frac{\dot{B}}{B}+\frac{\dot{C}}{C}\right).
\end{equation}

The Weyl tensor $(C_{\alpha\mu\beta\nu})$ can be broken into
electric and magnetic parts. However, the magnetic part vanishes for
plane symmetry, while the electric part is defined as
\begin{equation*}
E_{\alpha\beta}=C_{\alpha\mu\beta\nu}V^{\mu}V^{\nu},
\end{equation*}
whose non-vanishing components are
\begin{eqnarray*}
&&E_{11}=\frac{1}{3}B^{2}\varepsilon=E_{22},\quad
E_{33}=-\frac{2}{3}C ^{2} \varepsilon,
\end{eqnarray*}
where
\begin{eqnarray}\nonumber
&&\varepsilon=-\frac{1}{2A^{2}}\left[\frac{\ddot{C}}{C}-\frac{\ddot{B}}{B}
-\frac{\dot{A}\dot{C}}{AC}+\frac{\dot{B}^2}{B^2}+\frac{\dot{A}\dot{B}}{AB}
-\frac{\dot{B}\dot{C}}{BC}\right]\\\label{14}
&&-\frac{1}{2C^{2}}\left[\frac{A'C'}{AC}+\frac{A'B'}{AB}-\frac{B'^2}{B^2}
-\frac{B'C'}{BC}+\frac{B''}{B}-\frac{A''}{A}\right].
\end{eqnarray}
The electric part can also be written as
\begin{equation*}
E_{\alpha\beta}=\varepsilon(\chi_{\alpha}\chi_{\beta}-\frac{1}{3}h_{\alpha\beta}).
\end{equation*}

The mass function introduced by Taub \cite{14} in the presence of
electric charge is given as
\begin{equation}\label{15}
m(t,z)=\frac{B}{2}\left(\frac{\dot{B}^2}{A^2}
-\frac{B'^2}{C^2}\right)+\frac{{(s\mu_0)}^{2}}{2B}.
\end{equation}
Using the field equations (\ref{8}), (\ref{10}), (\ref{11}) and
(\ref{15}) in Eq.(\ref{14}), we obtain
\begin{equation}\label{16}
\varepsilon=4{\tilde{\mu}}{\pi}+4{\pi}(2{\eta}F-\Pi)+\frac{6{\mu_0^{2}s^{2}}}{2B^4}
-\frac{3m}{B^3},
\end{equation}
where $\Pi=\tilde{P_z}-{P_{\bot}},~\tilde{P_z}=P_z+\epsilon,~
\tilde{\mu}=\mu+\epsilon$. Differentiating Eq.(\ref{15}) with
respect to $z$ and then using the field equations (\ref{8}) and
(\ref{9}), it follows that
\begin{equation}\label{17}
m'=4\pi\left(\tilde{\mu}+\tilde{q}\frac{U}{E}\right)B'B^2+\frac{(\mu_{0}s)^{2}B'}{2B^{2}},
\end{equation}
where $U=\frac{\dot{B}}{A}$ is called the areal velocity of the
collapsing fluid, $\tilde{q}=q+\epsilon$ and $E=\frac{B'}{C}$.
Integrating this equation, after some simplification, we obtain
\begin{equation}\label{18}
\frac{3m}{B^3}=4{\pi}\tilde{\mu}-\frac{4\pi}{B^3}\int^z_{0}{B^3{\tilde{\mu'}}}dz
+\frac{4\pi}{B^3}\int^z_{0}{3B^2UC{\tilde{q}}}dz+\frac{3\mu_0^2}{B^3}\left[\frac{s^2}{2B}
+\int^z_{0}{\frac{B's^2}{2B^2}}dz\right].
\end{equation}
It provides a relationship between mass function and fluid
properties such as energy density, heat flux and electric charge.
Substituting this value in Eq.(\ref{16}), it follows that
\begin{eqnarray}\nonumber
\varepsilon=4{\pi}(2{\eta}F-\Pi)+\frac{3{\mu_0^{2}s^{2}}}{2B^4}
+\frac{4\pi}{B^3}\int^z_{0}{B^3{\tilde{\mu'}}}dz
&-&\frac{4\pi}{B^3}\int^z_{0}{3B^2UC{\tilde{q}}}dz\\\label{19}
&-&\frac{3\mu_0^2}{B^3}\int^z_{0}{\frac{B's^2}{2B^2}}dz.
\end{eqnarray}
This shows that the Weyl tensor depends on energy density
inhomogeneity and local anisotropy of pressure.

\section{Structure Scalars for the Charged Fluid}

In this section, we formulate structure scalars from the orthogonal
splitting of the Riemann tensor. For this purpose, we define the
following tensors
\begin{equation*}
Y_{\alpha\beta}=R_{\alpha\gamma\beta\delta}V^{\gamma}V^{\delta},\quad
X_{\alpha\beta}=^{*}R^{*}_{\alpha\gamma\beta\delta}V^{\gamma}V^{\delta}=
\frac{1}{2}\eta^{\varepsilon\rho}_{~~\alpha\gamma}
R^{*}_{\epsilon\rho\beta\delta}V^{\gamma}V^{\delta},
\end{equation*}
where $R^{*}_{\alpha\beta\gamma\delta}=
\frac{1}{2}\eta_{\varepsilon\rho\gamma\delta}R^{\epsilon\rho}_{\alpha\beta}$.
The tensors $Y_{\alpha\beta}$ and $X_{\alpha\beta}$ can be
decomposed in its trace and trace free components as
\begin{eqnarray}\nonumber
Y_{\alpha\beta}&=&\frac{1}{3}Y_{T}h_{\alpha\beta}+Y_{TF}\left(\chi_{\alpha}\chi_{\beta}
-\frac{1}{3}h_{\alpha\beta}\right),\\\label{21}X_{\alpha\beta}&=&\frac{1}{3}X_{T}h_{\alpha\beta}
+X_{TF}\left(\chi_{\alpha}\chi_{\beta}-\frac{1}{3}h_{\alpha\beta}\right).
\end{eqnarray}
Using the field equations (\ref{8}), (\ref{10}), (\ref{11}) and
(\ref{14}), the above components take the form
\begin{eqnarray}\nonumber
Y_{T}&=&4\pi\left(\tilde{\mu}+3\tilde{P_{z}}-2\Pi\right)+\frac{{\mu^2_{0}}s^{2}}{B^{4}},\\\nonumber
X_{T}&=&8\pi{\tilde{\mu}}+\frac{{\mu^2_{0}}s^{2}}{B^{4}},\\\nonumber
Y_{TF}&=&\varepsilon-4\pi\left(\Pi-2{\eta}F\right)+\frac{{\mu^2_{0}}s^{2}}{B^{4}},\\\label{22}
X_{TF}&=&-\varepsilon-4\pi\left(\Pi-2{\eta}F\right)+\frac{{\mu^2_{0}}s^{2}}{B^{4}}.
\end{eqnarray}
Substituting Eq.(\ref{19}) in (\ref{22}), it follows that
\begin{eqnarray}\nonumber
Y_{TF}&=&-8{\pi}{\Pi}+16{\eta}{\pi}F+\frac{5s^2\mu_{0}^{2}}{2B^4}
+\frac{4\pi}{B^3}\int^z_{0}{B^3\left(\tilde{\mu'}-\frac{3\tilde{q}UC}{B}\right)}dz\\\nonumber
&-&\frac{3\mu_0^2}{2B^3}\int^z_{0}{\frac{s^{2}B'}{B^2}}dz,\\\label{23}
X_{TF}&=&-\frac{s^2\mu_{0}^{2}}{2B^4}-\frac{4\pi}{B^3}\int^z_{0}{B^3\left(\tilde{\mu'}
+\frac{3\tilde{q}UC}{B}\right)}dz+\frac{3\mu_0^2}{2B^3}\int^z_{0}{\frac{s^{2}B'}{B^2}}dz,
\end{eqnarray}
which describes the density inhomogeneity and local anisotropy of
the fluid.

For the sake of convenience, we can introduce the following
effective variables as
\begin{eqnarray}\nonumber
-\left(T^{0}_{0}+S^{0}_{0}\right)&=&\mu_{eff}=\tilde{\mu}+\frac{\mu_{0}^{2}s^{2}}{8{\pi}B^{4}},\\\nonumber
\left(T^{1}_{1}+S^{1}_{1}\right)&=&P^{eff}_{z}=\left(\tilde{P_z}-\frac{4}{3}{\eta}F\right)-
\frac{\mu_{0}^{2}s^{2}}{8{\pi}B^{4}},\\\nonumber
\left(T^{2}_{2}+S^{2}_{2}\right)
&=&P^{eff}_{\bot}=\left(P_{\bot}+\frac{2}{3}{\eta}F\right)
+\frac{\mu_{0}^{2}s^{2}}{8{\pi}B^{4}},\\\nonumber
P^{eff}_{z}-P^{eff}_{\bot}&=&\Pi^{eff}=\left(\tilde{P_{z}}-P_{\bot}\right)
-2{\eta}{F}-\frac{\mu_{0}^{2}s^{2}}{8{\pi}B^{4}},\\\label{24}
\Pi^{eff}&=&\Pi-2{\eta}{F}-\frac{\mu_{0}^{2}s^{2}}{8{\pi}B^{4}}.
\end{eqnarray}
These equations show that the effective variables have the
resemblance with the ordinary variables from all the contributions
(viscosity and electric charge). In the light of above effective
variables, the structure scalars become
\begin{eqnarray}\nonumber
X_{T}&=&8\pi{\tilde{\mu}_{eff}},\\\nonumber
Y_{T}&=&4\pi\left(\tilde{\mu}_{eff}+3\tilde{P_{z}}-2\Pi^{eff}\right),\\\nonumber
Y_{TF}&=&-8\pi\Pi^{eff}+\frac{4\pi}{B^3}\int^z_{0}{B^3\left(\mu'_{eff}
-\frac{3\tilde{q}UC}{B}\right)}dz,\\\label{25}
X_{TF}&=&-\frac{4\pi}{B^3}\int^z_{0}{B^3\left(\tilde{{\mu}'}_{eff}-\frac{3\tilde{q}UC}{B}\right)}dz.
\end{eqnarray}
We see that the charge contribution is present in the effective
variables. In the absence of electric charge, the structure scalars
are directly obtained from the above equations by replacing the
effective variables with the ordinary ones.

In order to understand the physical significance of the electric
charge in structure scalars, we use Raychaudhuri equation which
gives the evolution of expansion and the shear. In the absence of
dissipation, $X_{TF}$ controls inhomogeneities in the energy
density. Moreover, it provides a differential equation which yields
the inhomogeneity factor and a relationship between the Weyl tensor
as well as other physical variables. The evolution equation for
expansion leads to
\begin{equation}\label{26}
-Y_{T}=V^{\alpha}\Theta_{;\alpha}+\frac{1}{3}{\Theta}^{2}
+{\sigma}^{\alpha\beta}{\sigma}_{\alpha\beta}-a^\alpha_{~;\alpha},
\end{equation}
The evolution equation for shear becomes
\begin{equation}\label{27}
Y_{TF}=\chi^\alpha a_{;\alpha}+a^{2}-\frac{aB'}{BC}-V^\alpha
F_{;\alpha}-\frac{2}{3}F{\Theta}-\frac{1}{3}F^{2}.
\end{equation}
We see that the evolution equations for expansion and shear are
independent of charge contribution. This implies that the electric
charge does not play any role in these two equations. Moreover,
these equations turn out to be the same as in spherically symmetric
spacetime. Finally, the differential equation for the Weyl tensor
and energy density inhomogeneity can be written as
\begin{equation}\label{28}
(X_{TF}+4\pi\mu_{eff})'=-X_{TF}\frac{3B'}{B}+4\pi\tilde{q}C(\Theta-F),
\end{equation}
which gives $X_{TF}$ as the inhomogeneity factor. This corresponds
to the charged spherically symmetric and also non-charged case by
replacing the effective energy density with the ordinary one.

\section{Structure Scalars for Dust with Cosmological Constant}

In the dust collapse, the role of density inhomogeneities \cite{15},
especially in the formation of naked singularities has been
extensively discussed in the literature. Eardley and Smarr
\cite{16a} discussed the inhomogeneous generalization of the
Oppenheimer-Snyder spherical dust collapse and found naked
singularities if the collapse is sufficiently inhomogeneous. Waugh
and Lake \cite{16b} found the necessary conditions for the formation
of the naked singularities. Joshi and Dwivedi \cite{16c}
investigated the occurrence and nature of a naked singularity for
the inhomogeneous gravitational collapse for spherical symmetry.
Moreover, they \cite{16d} discussed the structure of naked
singularities. The energy-density inhomogeneity also occurs due to
the presence of dissipation \cite{17}.

Here, we consider a special case of dust with non-vanishing
cosmological constant. Also, we take dissipation as well as shear
effects zero. In this case, the energy-momentum tensor takes the
form
\begin{equation}\label{29}
T_{\alpha\beta}=8{\pi}{\mu}V_{\alpha}V_{\beta},
\end{equation}
and the field equation has the form
\begin{equation}\label{30}
G_{\alpha\beta}=T_{\alpha\beta}-{\Lambda}g_{\alpha\beta},
\end{equation}
where $\Lambda$ is the cosmological constant. When $\Lambda$ is
negative, the universe shows contraction and inhomogeneity will
increase otherwise it will decrease. In comoving coordinate system,
the fluid represents geodesic behavior and $A'$ becomes zero.
Further, the re-scaling of the time coordinate gives $A=1$.
Consequently, the mass function reduces to
\begin{equation}\label{31}
m=4\pi\int^z_{0}{\mu}B'B^2dz+\frac{\Lambda}{6}B^3.
\end{equation}
After some manipulations, the mass function and $\varepsilon$ for
dust fluid with cosmological constant become
\begin{eqnarray}\nonumber
\frac{3m}{B^3}&=&4{\pi}{\mu}+\frac{\Lambda}{2}-\frac{4\pi}{B^3}\int^z_{0}{{\mu'}}dz,\\\label{32}
\varepsilon&=&\frac{4\pi}{B^3}\int^z_{0}{B^{3}{\mu'}}dz.
\end{eqnarray}
In this case, the scalar functions take the following form
\begin{equation}\label{33}
Y_{T}=4{\pi}{\mu}-\Lambda,\quad Y_{TF}=\varepsilon,\quad
X_{T}=8{\pi}{\mu}-\Lambda, \quad X_{TF}=\varepsilon.
\end{equation}
Also, the evolution equations for the expansion and shear become
\begin{eqnarray}\nonumber
V^\alpha\Theta_{;\alpha}+\frac{1}{3}\Theta^2+\frac{2}{3}F^2-a^\alpha_{;\alpha}&=&-4\pi\mu+\Lambda=-Y_T,\\\nonumber
-V^{\alpha}F_{;\alpha}-\frac{2}{3}\Theta
F-\frac{1}{3}F^{2}&=&\varepsilon=Y_{TF},
\end{eqnarray}
respectively. The differential equation for the inhomogeneity factor
can be written as
\begin{equation*}
(X_{TF}+4\pi\mu)'=-X_{TF}\frac{3B'}{B}.
\end{equation*}
It follows from here that $\mu'=0$ if and only if $X_{TF}=0$,
indicating $X_{TF}$ as the inhomogeneity factor.

\section{Summary}

We have investigated a set of scalar functions corresponding to the
charged plane symmetric distribution. We have found that $X_T$
corresponds to the energy density of the fluid with the contribution
of electric charge. In the absence of dissipation, $X_{TF}$ controls
the energy density inhomogeneity with the passage of time. Also,
$Y_T$ turns out to be the mass density while $Y_{TF}$ have the
interaction of both the energy density inhomogeneity and local
anisotropy. It is noted that the evolution of the shear and
expansion have the same contribution to $Y_T$ and $Y_{TF}$ as in the
spherically symmetric case. Also, it has the same effect on the
inhomogeneity factor. For the dust case with cosmological constant,
it is found that evolution of the expansion scalar is affected by
the term $\Lambda$ in $Y_T$, while evolution of the shear and
inhomogeneity factor remains the same. In this case, the relevant
factor for inhomogeneity is the Weyl tensor and stability of the
homogeneous  energy density is equivalent to the stability of the
conformal flatness.

\end{document}